\documentclass{article}
\usepackage[numbers,sort&compress]{natbib}
\usepackage{graphicx} 
\usepackage{tikz}
\usepackage{url}
\usepackage{booktabs}
\usetikzlibrary{arrows.meta, positioning, fit, backgrounds, shapes.geometric}
\title{From Skill Extraction to Multistakeholder Recommendation: A Two-Stage Framework for Bias Governance in Skills-Based Job Matching}

\author{
    Andrea Forster \and Gregor Autischer \and Dominik Kowald \and Simone Kopeinik\\
    Know Center Research GmbH, Graz, Austria
}

\date{July 2026}

\begin{document}

\maketitle
\begin{abstract}  
AI-based labor-market systems or platforms can affect access to job opportunities prior to organizational candidate rankings or hiring decisions. Such applications warrant caution, as biases in skill extraction, profile formation, and candidate-job matching may contribute to unfair treatment of candidates. In this paper, we propose a two-stage framework for detecting and governing bias in skills-based job matching. Stage 1, skill extraction and profile formation, addresses how candidates provide skills and preferences to the system, how the system extracts and structures this information, and the bias risks this entails, with a focus on chatbot-based elicitation. Stage 2, multistakeholder candidate-job recommendation, would embed this information in a recommender system in which candidate, company, and regulatory objectives are represented by separate agents, each producing an independent candidate-job ranking; these rankings would be combined through social choice-based aggregation into a single, auditable recommendation.
The two stages are connected by a shared distinction between hard constraints, which require correction before processing continues, and soft constraints, which are logged to inform later decisions. Following an AI Act-aligned assessment methodology (based on the Fraunhofer AI Assessment Catalog), we propose using distributional auditing and counterfactual testing to produce a Stage 1 bias inventory sorted into hard and soft constraints, with the latter informing fairness thresholds for Stage 2. The same logic would apply to Stage 2: fairness metrics crossing predefined thresholds would trigger an adapted recommendation process, while smaller deviations would be logged as bias reports and persistent fairness states.
The result is a framework linking certification-oriented assessment, technical bias detection, and stakeholder-aware recommendation into a single governance layer for labor-market AI under the EU AI Act.
\end{abstract}

\section{Introduction}
Automated skill extraction and job recommendation systems can shape access to labor-market opportunities by influencing how candidates are represented, which jobs they are shown, and which candidates businesses are encouraged to consider. Although skills-based matching can help connect job seekers with relevant opportunities, it can also reproduce inequalities contained in historical labor-market data. In particular, historical assignment patterns can restrict mobility between gender-typical occupations when they are incorporated into skills-based matching models~\cite{adhikari2024gender}. These effects may accumulate with repeated recommendations, meaning that fairness must be considered as a long-term property of the system rather than merely the quality of an individual recommendation~\cite{ge2021towards,scher2023modelling}. Bias can arise at multiple points in this process: candidate information may be collected or represented incorrectly, and matching may amplify those errors or reproduce existing inequalities.

This paper proposes a two-stage framework for detecting and governing bias in job platforms that use AI components for skill extraction, candidate profile building, and recommendation. Stage 1, skill extraction and profile formation, transforms candidate information into structured data. Stage 2, multistakeholder candidate-job recommendation, uses this structured information to produce and present job recommendations. The two stages are connected by a shared mechanism of hard and soft constraints, described below, which governs both the transition between stages and the presentation of individual recommendations.

Candidate information entering Stage 1 may be provided manually, extracted from a CV, or retrieved through a chatbot conversation or an interview transcription. In our discussion, we focus on chatbot-based ingestion because the involvement of a language model and its interaction with the user introduces additional risks related to language, communication style, and the extraction of incorrect or incomplete skills. For completeness, a corresponding ingestion process is also required for job postings provided by businesses; however, this paper focuses on the candidate side, treating job-posting ingestion as a possible extension of the same framework.

In Stage 2, the structured candidate profile is used for multistakeholder candidate-job recommendation. Rather than producing a single ranking directly, the framework represents Candidate, Company, and Regulator objectives through separate agents, each of which produces an independent ranking of candidate-job matches. These rankings are then combined through social-choice-based aggregation into a single, auditable recommendation, allowing the trade-offs between stakeholder objectives to remain visible rather than being resolved implicitly inside a single scoring function.

The transition between the two stages, and the presentation of each recommendation, are governed by structured bias assessments that classify findings as hard or soft constraints. Hard-constraint violations require correction or reprocessing before a profile or recommendation can proceed: examples include missing information, unresolved contradictions, or materially incorrect extractions in Stage 1, and fairness metrics crossing predefined thresholds in Stage 2. Soft constraints capture findings that are tolerable but should not be discarded, such as recurring differences in how skills are extracted across demographic groups or communication styles. Rather than blocking a profile, these are logged and carried forward to inform the fairness thresholds applied in Stage 2's aggregation process.

We discuss this framework in relation to the EU AI Act, and suggest the Fraunhofer AI Assessment Catalog as a practical assessment tool for defining assessment criteria, evidence requirements, tests, and mitigation measures across the system. Although this does not constitute formal certification, it can support auditability and preparation for AI Act compliance.

\noindent The paper makes three contributions:
{\setlength{\leftmargini}{1.5em}
\begin{itemize}
    \item We propose a two-stage framework for detecting and governing bias across the job-platform pipeline, connecting candidate and job-information extraction with job matching and recommendation through a shared hard/soft constraint mechanism.
    \item We introduce a multistakeholder recommendation architecture in which autonomous agents represent candidate, company, and regulatory objectives, and their rankings are combined through transparent social choice-based aggregation.
    \item We suggest examining the framework from an EU AI Act perspective and provide a glimpse on how a structured assessment approach, based on the Fraunhofer AI Assessment Catalog, can support system-wide auditability and AI Act compliance preparation without being treated as formal certification.
\end{itemize}
}

\section{Background}
Automated skill extraction and job matching can encode and amplify labor-market bias, while job recommendation itself must balance the differing objectives of candidates, companies, and regulators. Where such systems are classified as high-risk under the EU AI Act, they are additionally subject to formal conformity assessment. We review prior work on each of these concerns.

\subsection{Bias in Skill Extraction and Job Matching}
Automated systems for job matching can risk encoding and amplifying societal inequalities present in historical labor market data.
A central challenge is that protected attributes, such as gender, ethnic origin, religion, disability, and age, as defined under EU anti-discrimination law~\cite{council_eu_nondiscrimination} can be absorbed implicitly occupational stereotypes~\cite{escobedo2024making}.
Adhikari et al.~\cite{adhikari2024gender} study this dynamic directly in the context of labor markets, showing that skills-based matching models can restrict gender mobility when historical assignment patterns are left unaddressed. Their work motivates our focus on auditing the statistical distributions of skill assignments across demographic groups as a prerequisite for fair recommendation.
Scher et al.~\cite{scher2023modelling} further demonstrate that the fairness effects of data-driven job-seeker support are not static: targeted interventions can produce long-term distributional shifts that differ substantially from their short-term effects, underscoring the need for longitudinal monitoring. Fairness interventions in recommendations must therefore be understood as a long-term property~\cite{ge2021towards, scher2023modelling}. This can, for instance, be achieved through dynamic fairness-aware re-ranking as proposed in~\cite{aird2024dynamic}, and the use of beyond-accuracy metrics to capture distributional fairness over time.

\subsection{Multistakeholder Fairness in Recommender Systems}
\subsubsection{Stakeholder Objectives}
Our work builds on the growing literature of multistakeholder fairness in recommender systems, where algorithms must balance diverse and potentially competing priorities among stakeholders~\cite{burke2017multisided}. 
Job matching is inherently a multi-sided problem, involving job seekers, recruiters, companies, and regulatory or institutional third parties. Traditional recommender systems typically optimize for a single objective (typically user relevance). Each stakeholder, however, has distinct and often competing fairness objectives. Kaya and Bogers~\cite{kaya2025mapping} explicitly map this stakeholder landscape in the context of candidate recommendation for algorithmic hiring, identifying tensions between consumer-side fairness (equitable exposure for candidates) and provider-side fairness (relevant matches for recruiters), and between both of these and regulatory constraints such as proportional representation of protected groups. Their analysis informs our decision to operationalize each stakeholder perspective as a separate policy-based agent with its own fairness metrics. 

\subsubsection{Social Choice Aggregation}
Many fields (e.g., traditional multi-agent systems, group recommender systems) turn to computational social choice-based methods. This provides mathematically grounded, transparent, and equitable methods for aggregating individual preferences into collective decisions~\cite{dignum2025agentifying, popescu2013group}. Aird et al.~\cite{aird2024dynamic} operationalize computational social choice theory to integrate provider-side fairness concerns into recommendations on a peer-to-peer lending platform. They propose a framework in which distinct fairness concerns (e.g., targeted exposure for loans by a specific demographic group) are assigned to autonomous agents and dynamically integrated via formal allocation and aggregation mechanisms. This work inspires our multi-agent consensus architecture for the labor market. To achieve a principled consensus among these agents, the existing literature offers various voting rules for constructing a collective social welfare function~\cite{brandt2016handbook, aird2024dynamic}. Voting rules have different properties (axioms) that they fulfill, for instance, to prevent manipulation~\cite{brandt2016handbook}. The results can differ based on the chosen voting method. For instance, Condorcet-based methods favor majority-winning candidate-job matches, whereas Borda count aggregation tends to also favor more niche objectives~\cite{brandt2016handbook}. The outcomes of different voting mechanisms must be assessed and chosen based on the specific use case.

\subsection{Regulation and Assessment}
The use of machine learning and AI for skill extraction and job recommendation requires compliance with applicable regulatory obligations. The EU AI Act establishes the relevant legal framework, while structured assessment catalogs can help translate its requirements into concrete assessment criteria, and technical or organizational measures.

\subsubsection{EU AI Act and Labor Market AI}
The European Union's AI Act establishes a risk-based framework for AI systems that entered into force in 2024~\cite{european-union_regulation_2024,mueck_etsi_2022}. Under the AI Act, digital systems used for recruitment, candidate selection, or comparable employment-related decisions may be categorized as high-risk~\cite{pimentel_why_2024}, as they can affect access to employment opportunities~\cite{baum_fear_2023,cheong_transparency_2024}.

For high-risk systems, the AI Act sets out requirements that go beyond the general claims of fairness or reliability. Providers must address risk management, data governance, technical documentation, record keeping, transparency, human oversight, accuracy, robustness, and cybersecurity~\cite{european-union_regulation_2024,werry_eu_2024}. These obligations are the reference point for any AI assessment~\cite{autischer_towards_2026}.

\subsubsection{Assessment Catalogs and the Compliance Gap}
Structured assessment catalogs support AI Act preparation by translating broad trustworthiness requirements into concrete questions and evidence needs~\cite{poretschkin_ai_2023,autischer_ai_2025}. They must still be distinguished from formal conformity assessment. The AI Act also relies on harmonized standards to operationalize its requirements. Conformity with applicable harmonized standards can provide a presumption of conformity with the requirements covered by those standards. However, the relevant standards are not yet fully available~\cite{werry_eu_2024,kilian_european_2025}. Practical assessment catalogs were generally developed before the finalization of these standards or outside the formal standardization process and therefore do not, by themselves, establish conformity with the AI Act~\cite{poretschkin_ai_2023,autischer_practical_2025}.

This creates a possible compliance gap. An assessment catalog can structure documentation and tests, but it does not lead to an EU Declaration of Conformity or does not automatically show that a system is compliant with the AI Act~\cite{euaiact_key_2024,holistic_conformity_2023}. ISO/IEC 42001 and emerging standards can support governance, but do not remove product-level AI Act obligations~\cite{chatard_eu_2024,vanta_how_2025}.

\subsubsection{Fraunhofer AI Assessment Catalog as Practical Assessment Tool}
The Fraunhofer AI Assessment Catalog provides a practical assessment tool for structured assessments~\cite{autischer_ai_2025}. Its process starts with an AI Profile describing functionality, application context, operating environment, and technical structure~\cite{poretschkin_ai_2023}. It then uses life-cycle and protection-requirement analyses across dimensions such as fairness, autonomy and control, transparency, reliability, safety and security, and data protection~\cite{poretschkin_ai_2023}.

The catalog decomposes each dimension into risk objectives, criteria, measures, and an overall assessment. This is useful across all stages of our framework because it supports compliance preparation while also revealing technical errors, biases, and fairness risks that might otherwise remain undetected~\cite{sai-fi-de-nl-no-uk_auditing_2023,winter_trusted_2021}.

\subsubsection{Lessons Learned from Prior Certification Processes}
In prior use, the Fraunhofer Catalog proved to be effective in organizing an assessment, while on the down-side the process is documentation-heavy~\cite{autischer_ai_2025}. A catalog can reveal missing artifacts and unclear boundaries. Prior work has shown that technical testing during an auditing process can surface uncertainty, robustness, and subgroup performance issues that documentation-only approaches may miss~\cite{autischer_towards_2026}.

Assessment findings can also be translated into concrete development tasks. In prior work, identified gaps led to dataset curation, additional fairness and reliability metrics, robustness tests, architectural changes, and improved documentation~\cite{autischer_towards_2026}.  Certification should, therefore, also be considered as an opportunity to actively improve the system and its documentation.

\section{Framework Description}
\label{sec:framework}
\begin{figure}[p]
    \centering
    \includegraphics[width=0.95\linewidth]{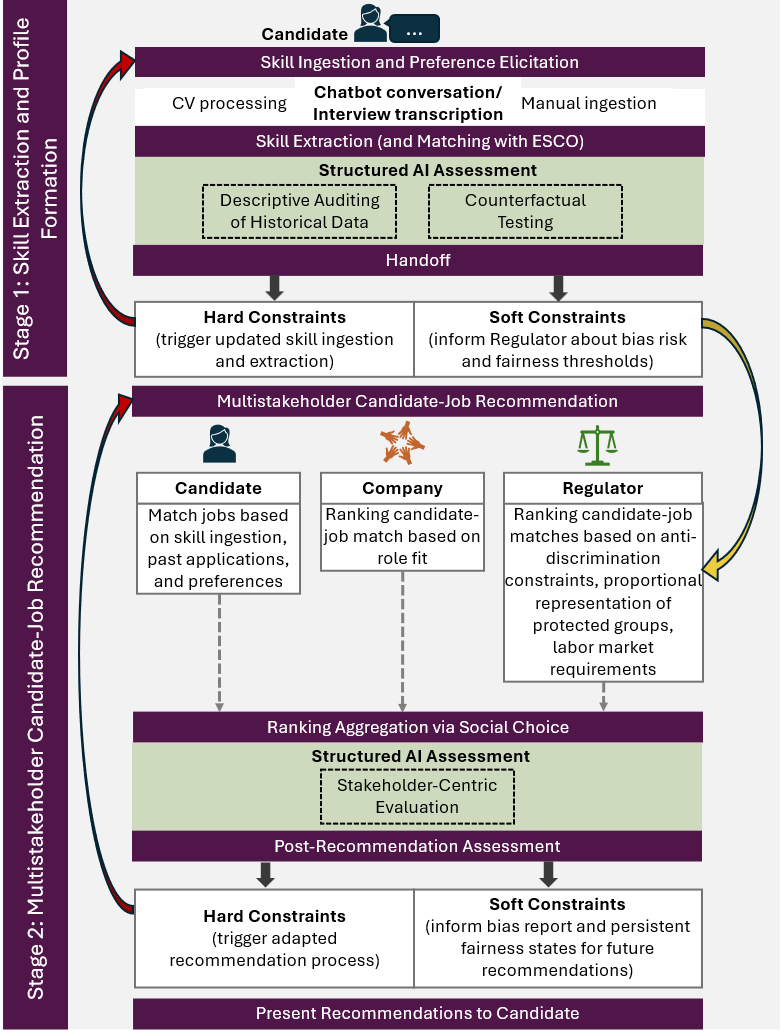}
    \caption{Two-stage conceptual framework for bias detection and governance in skills-based job matching: candidate skill extraction and assessment (Stage 1) hand off hard and soft constraints to a multistakeholder recommendation and post-recommendation governance process (Stage 2).}
    \label{fig:architecture}
\end{figure}

We propose a two-stage framework for bias detection and governance on skills-based job matching and recommendation platforms, where machine learning and AI are used across several stages of the pipeline (Figure~\ref{fig:architecture}). Throughout the framework, we distinguish hard constraints (violations that block downstream use of an output and require correction or reassessment) from soft constraints (residual risks that do not block processing but are logged and passed forward to inform decision-making). We apply this same hard/soft logic to two different objects across the two stages: profile validity in Stage 1, and fairness-metric thresholds on the aggregated ranking in Stage 2. This shared vocabulary reflects a shared governance mechanism, even though the underlying object of assessment differs.

In Stage 1, candidates and businesses provide information in unstructured form (e.g., via chatbot conversations or job postings), which is transformed into structured profiles. We briefly discuss all input paths in \ref{sec:stage1}, yet focus on chatbot-based candidate data ingestion, since its use of large language models introduces particularly complex interaction and extraction risks; the same assessment logic extends to business-side ingestion (e.g., job-posting processing), which we leave to future work. Bias detection methods, i.e., historical distributional auditing and counterfactual testing, both situated within a structured AI assessment, produce a bias inventory that flags hard and soft constraints. Hard-constraint violations require correction or review before a profile may proceed, while soft constraints are forwarded to the multistakeholder recommendation stage, where they inform the Regulator agent's objectives.

Stage 2 takes the structured candidate profiles and soft-constraint findings from Stage 1 as input to an agent-based ranking approach, in which candidate-job matches are ranked separately from the perspectives of the candidate, the company, and the regulatory stakeholder. These rankings are combined through social-choice-based aggregation into a single, auditable recommendation.

Stakeholder-centric evaluation metrics then guide a post-recommendation assessment that mirrors the Stage 1 logic: hard constraints, i.e., fairness metrics crossing predefined thresholds, re-trigger an adapted recommendation process in which different agent-allocation and aggregation strategies are tested to improve fairness and mitigate bias; soft constraints trigger a bias report, where the resulting fairness states of the stakeholder agents are stored.

We further propose structured AI assessment to the skill extraction and recommendation components as a regulatory layer, aligning the framework with requirements for high-risk applications under the EU AI Act. This supports scoring the system across life-cycle and protection-requirements analyses, which can in turn surface additional counterfactual tests or refined audit criteria to strengthen bias detection and mitigation across the pipeline.

\subsection{Stage 1: Skill Extraction and Profile Formation}
\label{sec:stage1}
The first stage looks at how information given by the candidate is converted into structured data to form a profile, i.e., skill extraction. Skills may be extracted via CV processing, chatbot or interview-based elicitation, or manual entry. Each method bears certain risks. CV processing must be checked for document layout, language, formatting, and cultural effects that can erase or distort education, experience, skills, certifications, gaps, or constraints. Chatbot- and interview-based elicitation adds interaction challenges because the system dynamically decides what to ask, when to stop, how to summarize, and which facts to translate into profile data. The option of manual entry gives candidates more direct control, but still requires validation, as users may omit skills, misunderstand categories, or enter inconsistent values. Across all options, the assessment process starts when candidate information is submitted and ends when structured profile data is stored. For theoretical grounding and standardization among disciplines, the extracted skills should be aligned with an established framework, such as provided by the European Skills, Competences and Occupations (ESCO)~\cite{esco2026} taxonomy. A central compliance consideration is whether the resulting profile is accurate, complete, reviewable, and robust to alterations of protected- or proxy-attributes.

\subsubsection{Bias Risks}
We narrow our focus to chatbot-based elicitation for the remainder of Stage 1's detailed discussion. Skill extraction from unstructured dialogs introduces risks beyond structured CV parsing. Candidates describe the same competence with different confidence, detail, vocabulary or language fluency. If the model rewards assertive self-presentation, it can under-extract skills from cautious users or users writing in a second language. There are many relevant risk factors, including dialect or gender-coded occupational language. The bias assessment should therefore test whether comparable profiles receive comparable follow-up questions, whether corrections are accepted, and whether non-standard skill evidence is converted into recognized profile fields.

\subsubsection{Bias Assessment Methods}
This section describes the methods used to construct the bias inventory that informs the hard and soft constraints applied at the handoff to the recommendation stage (Section~\ref{sec:handoff}). We address two distinct sources of bias: historical data bias, inherited from training and assignment patterns (distributional auditing), and model-internalized bias, arising from how the extraction model itself treats different linguistic presentations (counterfactual testing).

\paragraph{Distributional Auditing on Historical Data.}
\label{sec:auditing}
 
Machine learning and AI applications are typically trained on large-scale datasets, making them prone to learning the underlying biases present in those datasets~\cite{hort2024bias}. Consequently, it is essential to conduct a thorough analysis of training data to establish a ground truth of existing demographic disparities and to identify over- and underrepresentation. Such disparities often relate to protected attributes such as gender, ethnic origin, religion, disability, and age as defined under EU anti-discrimination law~\cite{council_eu_nondiscrimination}. Thus, as a first measure, we propose, for each skill or skill cluster, to record the share of assignments to each group and flag clusters where this share deviates substantially from the overall population distribution. 
Given the large number of competencies in a taxonomy such as the ESCO taxonomy, competencies can be grouped according to the hierarchical structure of the taxonomy. E.g., when gender distributions are projected onto the resulting clusters and compared to historical proportions over demographic data, visually distinct ``male-dominated'' and ``female-dominated'' regions indicate systematic assignment patterns. Clusters exhibiting high intra-cluster variance in gender distribution warrant further granular inspection. A distributional audit might, for instance, reveal that ``cleaning'' is assigned to female-coded profiles 87\% of the time in historical data, while ``electrical engineering'' is assigned to male-coded profiles 91\% of the time, even when both skills appear in the same interview transcript. These distributions become candidate thresholds for re-ranking. The objective is to balance highly skewed distributions over time using targeted re-ranking. We can further link skill clusters to labor market outcomes, estimating advertised positions and salary ranges (mean and variance) to identify which demographic groups are being steered toward lower-opportunity occupations. 

\paragraph{Counterfactual Testing}
\label{sec:counterfactual}
 
Distributional auditing establishes what patterns exist in historical data, whereas counterfactual testing probes whether the extraction model has internalized those patterns as learned associations. We propose two counterfactual procedures, both of which are recorded in the bias inventory produced by the structured assessment and inform the fairness constraints applied in Stage~2.
 
\emph{Gender expression substitution.} Within a held-out sample of interview transcripts, gender-coded linguistic expressions (e.g., gendered pronouns, stereotyped occupational terms) are systematically replaced with their counterparts. The skill extraction model is then rerun on the modified transcripts, and the resulting skill profiles are compared against the originals. A significant divergence in extracted skills (for instance, ``cleaning'' being assigned less frequently when a transcript is de-feminized) indicates that the model has encoded gender-specific associations beyond what the content of the description warrants~\cite{hort2024bias}.
 
\emph{Stereotype injection.} Stereotypical phrases are injected into otherwise neutral transcripts, and the resulting skill recommendations are compared to the neutral baseline. If injecting phrases associated with a particular demographic group systematically shifts the extracted skill profile toward that group's historical assignments, the model is propagating stereotypes rather than responding to substantive skill content~\cite{iso2025evaluating}.

\subsection{Handoff Phase bridging Skill-Extraction and Recommendations Stage}
\label{sec:handoff}
The output of the historical data audit and counterfactual tests, situated within Stage 1's structured AI assessment, forms a bias inventory that governs how a candidate profile moves into Stage 2. Based on extraction failure modes, uncertain skills, missing fields, and the bias-auditing results described above, the inventory distinguishes between hard constraints, which gate whether a profile may proceed at all, and soft constraints, which are carried forward as fairness thresholds for the Regulator agent.

\subsubsection{Hard Constraints}
Hard constraints correspond to unresolved extraction failures or high-stakes fairness issues. This includes missing required fields, unresolved contradictions, or an inferred skill profile that clearly misrepresents a candidate in a way tied to a protected characteristic. For these cases, the system triggers a correction or re-elicitation step and only proceeds once the issue has been addressed. Allowing a misrepresented profile to proceed would expose that specific candidate to direct discrimination risks. Therefore, before a profile can reach the recommendation stage, it receives one of three inventory outcomes: \emph{approved}, \emph{requires correction}, or \emph{blocked pending review}.

\subsubsection{Soft Constraints}
Once a profile is approved, findings that do not warrant blocking are logged as soft constraints and passed to the Regulator agent. This includes the identified at-risk skill clusters, affected demographic groups, and fairness thresholds derived from the historical bias audit and counterfactual tests, as well as systematic patterns such as recurring diversity bias in inferred skill profiles across candidates. These findings inform the Regulator agent's fairness thresholds in Stage 2 (Section~\ref{sec:agents}). 

\subsection{Stage 2: Multistakeholder Candidate-Job Recommendation}
\label{sec:stage2}
The second stage of the framework comprises a multistakeholder recommender system for candidate-job matching. The recommender system is a multi-agent system where each agent represents a specific stakeholder. These agents can be implemented either as traditional recommender algorithms, deterministic re-ranking and filtering strategies, or language models guided by natural language instructions to produce a ranking. Each agent should produce its individual ranking of candidate-job matches based on its own policy to subsequently allow for transparent aggregation~\cite{forster2026fair}. For this framework, a matching algorithm compares the extracted skills and stated preferences of users (Stage 1) against relevant open positions. To aggregate these individual rankings into a final candidate-job match, we draw on voting methods from computational social choice theory, which combine the rankings of all instantiated agents into a single consensus ranking.

\subsubsection{Two-Sided Matching}
Labor market recommender systems are inherently two-sided, in contrast to domains such as movie or product recommendation, where the primary objective is to match a single user to a satisfying set of items. In the labor market context, job seekers aim to identify relevant and attainable positions, while recruiters and employers aim to identify suitable applicants from a pool of candidates~\cite{rus2026joint}. Fairness concerns, therefore, arise on both sides, with job seekers potentially exposed to biased or discriminatory treatment in the opportunities presented to them, while businesses may be concerned with the representativeness and quality of the candidate pools they receive.

\subsubsection{Stakeholder Agents}
\label{sec:agents}
Our framework integrates several stakeholder perspectives and operationalizes them as separate agents. These can be based on large language models or deterministic re-ranking functions that take in the original matches based on the skill extraction. In this case, we illustrate this using three agents: \emph{Candidate}, \emph{Company}, and \emph{Regulator}. However, it is possible to extend this with additional, potentially more fine-grained agents. 

\paragraph{Candidate.}
This agent represents the job seeker. This agent's rankings can, for instance, be created by a baseline recommender system that matches the extracted skills from Stage 1, as well as stated user preferences and information about past applications with relevant and available jobs. It provides a ranking of candidate-job matches based on skills, preferences, location, and work-mode constraints against available postings.

\paragraph{Company.} 
This agent represents the recruiter or company that is hiring and their interest in identifying strong matches. This agent can, for instance, re-rank and filter candidate-job matches for a given role according to role fit (i.e., how well a candidate's qualifications, experience, and stated preferences align with the job's requirements).

\paragraph{Regulator.}
This agent represents broader societal and regulatory interests in the labor market. It functions as a bias regulator to increase the overall fairness of the system. This agent re-ranks or constrains candidate-job matches according to the previously defined fairness thresholds for skill clusters and demographic groups that are at risk of discrimination in the labor market. It can keep track of fairness states for different demographic groups, thereby steering the recommendations by filtering and re-ordering candidate-job matches to ensure proportional representation of protected groups. 

\subsubsection{Aggregation and Evaluation}
This section covers how individual stakeholder rankings are combined into a single recommendation, and how the resulting recommendation is evaluated across stakeholder objectives.

\paragraph{Fair Aggregation of Diverse Stakeholder Agents.}
Each stakeholder agent generates candidate-job rankings based on its own objectives and policies. To aggregate these individual rankings into a unified, stakeholder-aware ranking, we draw on computational social choice theory. Different voting methods should be compared, since the choice of method can lead to different outcomes. The most suitable aggregation method is highly dependent on the specific use case. 

\paragraph{Stakeholder-Centric Evaluation.}
\label{sec:eval}
At the multi-agent level, the aggregated top-k recommendations can be assessed using standard accuracy metrics (e.g., nDCG, recall) alongside diversity metrics that capture item- and group-level fairness, such as the Gini index or normalized entropy~\cite{deldjoo2025understanding}. At the individual level, fairness can be evaluated per stakeholder objective. For the candidate side, demographic parity can be assessed by comparing the demographic distribution of recommended jobs against the historical baseline, for instance via Kullback-Leibler or Jensen-Shannon divergence~\cite{lesota2021analyzing}, while popularity bias can be tracked using measures such as popularity lift~\cite{abdollahpouri2019impact}. Aird et al.~\cite{aird2024dynamic} further propose fairness regret as the gap between an agent's ranking under perfect adherence to its own fairness objective and the ranking actually produced by the aggregated system. Beyond these offline metrics, longer-term policy drift across repeated interactions should also be monitored, and offline results should be complemented with online studies to capture the stakeholders' perceived fairness and satisfaction with the recommendations.\\
Together, aggregation design and stakeholder-centric evaluation form the governance layer that determines whether a recommendation is released or triggers the adapted process described in Section~\ref{sec:post-rec}.

\subsection{Post-Recommendation Assessment}
\label{sec:post-rec}
After aggregation, the stakeholder-centric evaluation metrics (Section~\ref{sec:eval}) are compared against predefined fairness thresholds, following the same hard/soft constraint logic introduced in Stage 1 (Section~\ref{sec:handoff}). Hard-constraint violations occur when a fairness metric crosses a critical threshold for a stakeholder or protected group. Such violations re-trigger an adapted recommendation process: alternative agent configurations (e.g., adjusted regulator weighting or filtering strategy) and aggregation rules (e.g., switching voting methods) are tested, and the resulting ranking is re-evaluated against the same metrics before release.

Soft-constraint findings, such as deviations that remain within an acceptable but non-negligible range, do not block release but are logged in a bias report and appended to the fairness state maintained for each stakeholder agent. This state accumulates across recommendation cycles, allowing the Regulator agent to detect gradual drift that would not be visible in any single evaluation, and to inform threshold adjustments in future assessment iterations.

\subsection{AI Act Conformity Assessment}

The Fraunhofer AI Assessment Catalog could be applied to our framework as a structured methodology for identifying risks and documenting evidence. The assessment should cover the complete application, as well as isolated AI components and their interaction~\cite{autischer_ai_2025}.

The process begins with an AI Profile that defines the intended use, actors, inputs, outputs, interfaces, operating environment, human involvement, and system boundaries. A lifecycle and protection-requirement analysis then considers the relevance of fairness, autonomy and control, transparency, reliability, safety, security, and data protection~\cite{poretschkin_ai_2023}. Based on this analysis, system-specific criteria, measures, and evidence requirements are defined. This assessment logic is component-general: for Stage 1, the evidence requirements identified through lifecycle analysis can be met through the distributional auditing and counterfactual testing procedures described in Section~\ref{sec:auditing}. They can draw on evidence such as annotated CVs and conversations, ground-truth profiles, stored profile data, correction tests, and subgroup comparisons to reveal whether extraction quality differs across demographic groups, languages, communication styles, or career paths. For Stage 2, they can be met through the stakeholder-centric evaluation described in Section~\ref{sec:eval}, covering the matching process, the objectives and behavior of each stakeholder agent, and the aggregation mechanism. Here, evidence includes recommendation accuracy and stability, differences in exposure across groups, and scenarios in which agents produce conflicting rankings; the aggregation rule, agent weights, tie handling, and fairness objectives must be documented and evaluated accordingly.

The assessment should follow an iterative cycle in which identified deficiencies become development tasks and the modified system is reassessed using the same criteria. Previous work showed that this process can guide concrete technical improvements and that technical testing can reveal reliability and fairness problems that documentation or aggregate performance measures alone may overlook~\cite{autischer_towards_2026}.

Applying the catalog in this way could improve auditability and support preparation for AI Act compliance. However, the catalog is not a harmonized standard or a formal conformity-assessment procedure and therefore cannot, by itself, establish compliance with the AI Act~\cite{european-union_regulation_2024,autischer_towards_2026}.

\section{Discussion}
The central argument of this paper is that bias in AI-supported job platforms cannot be addressed only during candidate-job ranking. Recommendations also depend on how candidate and job information is collected and structured. Missing or incorrectly extracted skills, experience, preferences, or job requirements can affect the matches generated by the recommender system. Historical distributions and counterfactual tests, as part of a structured AI assessment, can help surface such patterns.

The distinction between hard and soft constraints governs the transition between Stages 1 and 2, either re-triggering elicitation or informing the Regulator agent's fairness objectives. The precise objectives, thresholds, and permissible interventions of the handoff remain design decisions requiring empirical validation. Strict constraints may unnecessarily prevent candidates from receiving recommendations, whereas weak constraints may allow unreliable profiles to proceed. 

Stakeholder concerns (Candidate, Company, and Regulator) are integrated into recommendations to make conflicting objectives explicit. While the Regulator agent inherits these fairness objectives, it should not alter demographic distributions by recommending positions that conflict with a candidate's qualifications, preferences, or constraints. Instead, it weighs fairness risks within the set of relevant and attainable matches, e.g., by re-ranking recommendations that are personalized for the candidate.

Individual rankings are aggregated using social choice methods. Aggregation does not itself guarantee fairness, since different voting rules, weights, and tie-breaking procedures can yield different outcomes. The post-recommendation assessment addresses this by re-triggering tests of alternative configurations whenever fairness metrics cross set thresholds, rather than treating the first aggregated ranking as final. These design choices must be documented and evaluated against recommendation relevance, candidate/company exposure, and each agent's assigned objectives.

The Fraunhofer AI Assessment Catalog provides a structured methodology for examining these risks across the framework. Stage 1 requires evidence concerning extraction quality, profile completeness, correction mechanisms, subgroup differences, and counterfactual stability. Stage 2 requires evidence concerning matching quality, agent behavior, group exposure, aggregation outcomes, and changes over time. Documentation should be complemented by technical testing, since previous certification work showed that aggregate performance measures may conceal reliability and subgroup-level problems. Applying the catalog can consequently improve auditability and compliance preparation, but it does not constitute a formal AI Act conformity assessment. The legal classification and applicable obligations must be determined for the specific implementation and intended use.

The framework remains conceptual and has not been evaluated as a complete system. Its detailed analysis focuses on chatbot-based candidate ingestion, while manual entry, CV processing, and company-side job postings require further examination. The framework is agnostic to fairness definitions, constraint thresholds, agent weightings, or voting rules. Future work should therefore implement the complete architecture, compare alternative configurations, and evaluate recommendation quality and fairness over repeated interactions.

\section{Conclusion}
This paper proposes a two-stage framework for detecting and governing bias in AI-supported job platforms, spanning skill extraction and profile formation, as well as multistakeholder candidate-job recommendation. Both stages are governed by the same hard/soft constraint logic: hard constraints block progression and require correction, while soft constraints are logged and carried forward to inform downstream fairness objectives. We illustrate this in detail for chatbot-based candidate ingestion (Stage 1) and its handoff to a multistakeholder recommender system in which candidate, company, and regulatory objectives are represented by separate agents and combined via social-choice-based aggregation (Stage 2), where the same constraint logic governs whether a recommendation is released or reprocessed. The Fraunhofer AI Assessment Catalog structures this process across both stages, supporting auditability and AI Act compliance preparation without constituting a formal conformity assessment.

The framework is conceptual and requires empirical validation. Future work should evaluate the remaining skill extraction and profile formation paths and business-side job postings, compare alternative constraint thresholds, agent configurations, and aggregation rules, and examine recommendation quality and fairness over repeated interactions, including appropriate human-review, monitoring, and data-protection procedures for a concrete implementation.

\bibliography{main.bib}

@misc{council_eu_nondiscrimination,
  author       = {{Council of the European Union}},
  title        = {Non-discrimination},
  howpublished = {Consilium},
  note         = {Retrieved April 21, 2026, from \url{https://www.consilium.europa.eu/en/policies/non-discrimination/}},
  year         = {2025}
}

@article{hort2024bias,
  title={Bias mitigation for machine learning classifiers: A comprehensive survey},
  author={Hort, Max and Chen, Zhenpeng and Zhang, Jie M and Harman, Mark and Sarro, Federica},
  journal={ACM Journal on Responsible Computing},
  volume={1},
  number={2},
  pages={1--52},
  year={2024},
  publisher={ACM New York, NY}
}

@inproceedings{escobedo2024making,
  title={Making alice appear like bob: A probabilistic preference obfuscation method for implicit feedback recommendation models},
  author={Escobedo, Gustavo and Moscati, Marta and Muellner, Peter and Kopeinik, Simone and Kowald, Dominik and Lex, Elisabeth and Schedl, Markus},
  booktitle={Joint European Conference on Machine Learning and Knowledge Discovery in Databases},
  pages={349--365},
  year={2024},
  organization={Springer}
}

@article{adhikari2024gender,
  title={Gender mobility in the labor market with skills-based matching models},
  author={Adhikari, Ajaya and Vethman, Steven and Vos, Daan and Lenz, Marc and Cocu, Ioana and Tolios, Ioannis and Veenman, Cor J},
  journal={AI and Ethics},
  volume={4},
  number={1},
  pages={163--167},
  year={2024},
  publisher={Springer}
}

@inproceedings{ge2021towards,
  title={Towards long-term fairness in recommendation},
  author={Ge, Yingqiang and Liu, Shuchang and Gao, Ruoyuan and Xian, Yikun and Li, Yunqi and Zhao, Xiangyu and Pei, Changhua and Sun, Fei and Ge, Junfeng and Ou, Wenwu and others},
  booktitle={Proceedings of the 14th ACM international conference on web search and data mining},
  pages={445--453},
  year={2021}
}

@inproceedings{kaya2025mapping,
  title={Mapping Stakeholder Needs to Multi-Sided Fairness in Candidate Recommendation for Algorithmic Hiring},
  author={Kaya, Mesut and Bogers, Toine},
  booktitle={Proceedings of RecSys'25},
  pages={257--267},
  year={2025}
}

@article{scher2023modelling,
  title={Modelling the long-term fairness dynamics of data-driven targeted help on job seekers},
  author={Scher, Sebastian and Kopeinik, Simone and Tr{\"u}gler, Andreas and Kowald, Dominik},
  journal={Scientific Reports},
  volume={13},
  number={1},
  pages={1727},
  year={2023},
  publisher={Nature Publishing Group UK London}
}

@article{aird2024dynamic,
  title   = {Dynamic Fairness-aware Recommendation through Multi-agent Social Choice},
  author  = {Aird, Amanda and Farastu, Paresha and Sun, Joshua and Stefancova, Elena and All, Cassidy and Voida, Amy and Mattei, Nicholas and Burke, Robin},
  journal = {ACM TORS},
  volume  = {3},
  number  = {2},
  pages   = {1--35},
  year    = {2024}
}

@article{dignum2025agentifying,
  title={Agentifying Agentic AI},
  author={Dignum, Virginia and Dignum, Frank},
  journal={arXiv preprint arXiv:2511.17332},
  year={2025}
}

@inproceedings{popescu2013group,
  title={Group recommender systems as a voting problem},
  author={Popescu, George},
  booktitle={International Conference on Online Communities and Social Computing},
  pages={412--421},
  year={2013},
  organization={Springer}
}

@book{brandt2016handbook,
  title={Handbook of computational social choice},
  author={Brandt, Felix and Conitzer, Vincent and Endriss, Ulle and Lang, J{\'e}r{\^o}me and Procaccia, Ariel D},
  year={2016},
  publisher={Cambridge University Press}
}

@inproceedings{rus2026joint,
  title={Joint Modeling of Candidate and Recruiter Preferences for Fair Two-Sided Job Matching},
  author={Rus, Clara and Mansoury, Masoud and Yates, Andrew and de Rijke, Maarten},
  booktitle={European Conference on Information Retrieval},
  pages={335--351},
  year={2026},
  organization={Springer}
}

@inproceedings{lesota2021analyzing,
  title={Analyzing item popularity bias of music recommender systems: are different genders equally affected?},
  author={Lesota, Oleg and Melchiorre, Alessandro and Rekabsaz, Navid and Brandl, Stefan and Kowald, Dominik and Lex, Elisabeth and Schedl, Markus},
  booktitle={Proceedings of RecSys'21},
  pages={601--606},
  year={2021}
}

@article{abdollahpouri2019impact,
  title={The impact of popularity bias on fairness and calibration in recommendation},
  author={Abdollahpouri, Himan and Mansoury, Masoud and Burke, Robin and Mobasher, Bamshad},
  journal={arXiv preprint arXiv:1910.05755},
  year={2019}
}

@inproceedings{autischer_ai_2025,
  title={{AI} {Certification} and {Assessment} {Catalogues}: {Practical} {Use} and {Challenges} in the {Context} of the {European} {AI} {Act}},
  shorttitle={{AI} {Certification} and {Assessment} {Catalogues}},
  url={https://proceedings.mlr.press/v294/autischer25a.html},
  booktitle={Proceedings of {Fourth} {European} {Workshop} on {Algorithmic} {Fairness}},
  publisher={PMLR},
  author={Autischer, Gregor and Waxnegger, Kerstin and Kowald, Dominik},
  month=jul,
  year={2025},
  note={ISSN: 2640-3498},
  pages={492--498}
}

@inproceedings{autischer_towards_2026,
  title={Towards {EU} {AI} {Act} {Compliance}: {Self}-{Certification} and {Fairness} {Alignment} for {Facial} {Emotion} {Recognition}},
  author={Autischer, Gregor and Waxnegger, Kerstin and Kowald, Dominik},
  booktitle={Proceedings of {Fifth} {European} {Conference} on {Algorithmic} {Fairness}},
  publisher={PMLR},
  year={2026},
  month=sep,
  note={ECAF'26}
}

@misc{european-union_regulation_2024,
  title={Regulation ({EU}) 2024/1689 of the {European} {Parliament} and of the {Council} of 13 {June} 2024 laying down harmonised rules on artificial intelligence ({Artificial} {Intelligence} {Act})},
  url={http://data.europa.eu/eli/reg/2024/1689/oj/eng},
  author={{European Union}},
  year={2024},
  month=jun,
  note={Official Journal of the European Union}
}

@techreport{poretschkin_ai_2023,
  title={{AI} {Assessment} {Catalog}},
  institution={Fraunhofer IAIS},
  author={Poretschkin, Maximilian and Schmitz, Anna and Akila, Maram and Adilova, Linara and Becker, Daniel and Cremers, Armin B. and Hecker, Dirk and Houben, Sebastian and Rosenzweig, Julia and Sicking, Joachim and Schulz, Elena and Voss, Angelika and Wrobel, Stefan},
  month=feb,
  year={2023}
}

@techreport{sai-fi-de-nl-no-uk_auditing_2023,
  title={Auditing machine learning algorithms},
  institution={Supreme Audit Institutions FI, DE, NL, NO, UK},
  author={{SAI-FI-DE-NL-NO-UK}},
  month=feb,
  year={2023}
}

@article{winter_trusted_2021,
  title={Trusted {Artificial} {Intelligence}: {Towards} {Certification} of {Machine} {Learning} {Applications}},
  author={Winter, Philip Matthias and Eder, Sebastian and Weissenböck, Johannes and Schwald, Christoph and Doms, Thomas and Vogt, Tom and Hochreiter, Sepp and Nessler, Bernhard},
  journal={arXiv preprint arXiv:2103.16910},
  year={2021},
  doi={10.48550/arXiv.2103.16910},
  url={http://arxiv.org/abs/2103.16910}
}

@techreport{mueck_etsi_2022,
  title={{ETSI} Activities in the field of Artificial Intelligence: Preparing the implementation of the European {AI} Act},
  author={Mueck, Markus and Cadzow, Scott and Wood, Suno},
  institution={ETSI},
  year={2022}
}

@misc{pimentel_why_2024,
  title={Why {AI} still needs regulation despite impact},
  author={Pimentel, Brandon},
  journal={Thomson Reuters Law Blog},
  url={https://legal.thomsonreuters.com/blog/why-ai-still-needs-regulation-despite-impact/},
  year={2024},
  month=feb,
  urldate={2024-10-28}
}

@article{baum_fear_2023,
  title={From fear to action: {AI} governance and opportunities for all},
  author={Baum, Kevin and Bryson, Joanna and Dignum, Frank and Dignum, Virginia and Grobelnik, Marko and Hoos, Holger and Irgens, Morten and Lukowicz, Paul and Muller, Catelijne and Rossi, Francesca and Shawe-Taylor, John and Theodorou, Andreas and Vinuesa, Ricardo},
  journal={Frontiers in Computer Science},
  volume={5},
  year={2023},
  doi={10.3389/fcomp.2023.1210421}
}

@article{cheong_transparency_2024,
  title={Transparency and accountability in {AI} systems: safeguarding wellbeing in the age of algorithmic decision-making},
  author={Cheong, Ben Chester},
  journal={Frontiers in Human Dynamics},
  volume={6},
  year={2024},
  doi={10.3389/fhumd.2024.1421273}
}

@misc{autischer_practical_2025,
  title={Practical Application and Limitations of {AI} Certification Catalogues in the Light of the {AI} Act},
  author={Autischer, Gregor and Waxnegger, Kerstin and Kowald, Dominik},
  year={2025},
  month=feb,
  publisher={arXiv},
  doi={10.48550/arXiv.2502.10398},
  url={http://arxiv.org/abs/2502.10398},
  note={arXiv:2502.10398}
}

@misc{werry_eu_2024,
  title={{EU} Standardization Supporting the Artificial Intelligence Act},
  author={Werry, Susanne and Ridgway, Simon and Kerr-Shaw, Simon and Silverstein, Michael},
  url={https://www.skadden.com/insights/publications/2024/10/eu-standardization-supporting-the-artificial-intelligence-act},
  year={2024},
  urldate={2025-11-17}
}

@article{kilian_european_2025,
  title={European {AI} Standards: Technical Standardization and Implementation Challenges under the {EU} {AI} Act},
  author={Kilian, Robert and Jack, Linda and Ebel, Dominik},
  journal={European Journal of Risk Regulation},
  volume={16},
  number={3},
  pages={1038--1062},
  year={2025},
  doi={10.1017/err.2025.10032}
}

@misc{euaiact_key_2024,
  title={Key Issue 2: Conformity Assessment and Self-Assessment},
  author={{EU AI Act}},
  url={https://www.euaiact.com/key-issue/2},
  year={2024},
  urldate={2025-11-28}
}

@misc{holistic_conformity_2023,
  title={Conformity Assessments in the {EU} {AI} Act: What You Need to Know},
  author={{Holistic AI}},
  url={https://www.holisticai.com/blog/conformity-assessments-in-the-eu-ai-act},
  year={2023},
  urldate={2025-11-28}
}

@misc{chatard_eu_2024,
  title={{EU} {AI} Act unpacked \#10: {ISO} 42001 - a tool to achieve {AI} Act compliance?},
  author={Chatard, Yannick and Riede, Lutz and Werkmeister, Christoph and Ehlen, Theresa and Roos, Philipp and Kirchmair, Verena and Knoke, Laura},
  url={https://technologyquotient.freshfields.com/post/102jcog/eu-ai-act-unpacked-10-iso-42001-a-tool-to-achieve-ai-act-compliance},
  year={2024},
  month=jul,
  urldate={2025-11-28}
}

@misc{vanta_how_2025,
  title={How {ISO} 42001 helps with {EU} {AI} Act compliance},
  author={{Vanta}},
  url={https://www.vanta.com/resources/iso-42001-and-eu-ai-act},
  year={2025},
  urldate={2025-11-28}
}

@misc{esco2026,
  author       = {{European Commission}},
  title        = {{ESCO (European Skills, Competences, Qualifications and Occupations) portal}},
  year         = {2026},
  url          = {https://esco.ec.europa.eu/},
  note         = {Accessed: 2026-07-08}
}

@article{forster2026fair,
  title={Fair Agents: Balancing Multistakeholder Alignment in Multi-Agent Personalization Systems},
  author={Forster, Andrea and Müllner, Peter and Helic, Denis and Lex, Elisabeth and Kowald, Dominik},
  journal={arXiv preprint arXiv:2605.02379},
  year={2026}
}

@article{deldjoo2025understanding,
  title={Understanding biases in ChatGPT-based recommender systems: Provider fairness, temporal stability, and recency},
  author={Deldjoo, Yashar},
  journal={ACM TORS},
  volume={4},
  number={2},
  pages={1--35},
  year={2025},
  publisher={ACM New York, NY}
}

@inproceedings{iso2025evaluating,
  title={Evaluating bias in LLMs for job-resume matching: Gender, race, and education},
  author={Iso, Hayate and Pezeshkpour, Pouya and Bhutani, Nikita and Hruschka, Estevam},
  booktitle={Proceedings of the 2025 Conference of the Nations of the Americas Chapter of the Association for Computational Linguistics: Human Language Technologies (Volume 3: Industry Track)},
  pages={672--683},
  year={2025}
}

@article{burke2017multisided,
  title   = {Multisided Fairness for Recommendation},
  author  = {Burke, Robin},
  journal = {Workshop on Fairness, Accountability, and Transparency in Machine Learning},
  year    = {2017}
}

\end{document}